\documentclass[journal]{IEEEtran}
\usepackage{amsmath, amssymb,amsthm}
\usepackage{graphicx, float, adjustbox, multirow, array}
\usepackage[caption=false, font=footnotesize, labelfont=sf,textfont=sf]{subfig}
\usepackage{authblk}
\usepackage{enumerate,enumitem, hyperref, siunitx}
\usepackage[numbers,sort,square,compress]{natbib}

\usepackage{cuted}
\usepackage{algorithm,algpseudocode}

\newcommand{\MB}[1]{\mathbb{#1}}

\newcommand{\MC}[1]{\mathcal{#1}}

\newcommand{\e}[1]{\mathrm{e}^{#1}}
\newcommand{\dif}{\mathrm{d}}
\newcommand{\del}{\partial}

\newcommand{\ceil}[1]{\left\lceil #1 \right\rceil}
\newcommand{\floor}[1]{\left\lfloor #1 \right\rfloor}
\newcommand{\imgwidth}{0.72\columnwidth}
\newcommand{\twidth}{0.47\columnwidth}

\renewcommand{\vec}[1]{{\mathbf{#1}}}
\newcommand{\vecsym}[1]{{\boldsymbol{#1}}}

\def\db{\si{\deci\bel}}
\title{Robust Direction-of-Arrival Estimation using Array Feedback
  Beamforming in Low SNR Scenarios}
\author{Parth Mehta$^{\dagger}$, Kumar Appaiah~\IEEEmembership{Member,~IEEE}, and Rajbabu Velmurugan~\IEEEmembership{Member,~IEEE}
\thanks{~}\thanks{Manuscript received \today.}}
\begin{document}
\maketitle
\begin{abstract}
A new spatial IIR beamformer based direction-of-arrival (DoA) estimation
method is proposed in this paper. We propose a retransmission based
spatial feedback method for an array of transmit and receive antennas
that improves the performance parameters of a beamformer viz. half-power
beamwidth (HPBW), side-lobe suppression, and directivity.
Through quantitative comparison we show that our approach
outperforms the previous feedback beamforming approach with a single 
transmit antenna, and the conventional beamformer. We then incorporate 
a retransmission based minimum variance distortionless response (MVDR) 
beamformer with the feedback beamforming setup. We propose two approaches, 
show that one 
approach is superior in terms of lower estimation error, and use that 
as the DoA estimation method. We then compare this approach with Multiple
Signal Classification (MUSIC), Estimation of Parameters using Rotation
Invariant Technique (ESPRIT), robust MVDR, nested-array MVDR, and reduced-dimension
MVDR methods. The results show that at SNR levels of 
-60~\db\ to -10~\db, the angle estiation error of the proposed method
is $20^{\circ}$ less compared to that of prior methods.
\end{abstract}
%
\IEEEpeerreviewmaketitle
%
\section{Introduction}
\label{sec:intro}
Beamforming is a well-studied concept in the field of spatial signal
processing using antenna arrays. Spatial beamforming, along with
direction estimation of the incoming signals and interference
mitigation has received significant attention lately, especially for
cognitive radar~\cite{haykin2006cognitive} systems, 
and systems wherein radar and communication need to
coexist~\cite{li2014robust,liu2017robust}. Such coexistence
necessitates robust beamforming with strong interference suppression,
and accurate direction-of-arrival (DoA) estimation capabilities. The
MVDR~\cite{van2002optimum}
beamformer is one candidate that offers simultaneous beamforming and
interference suppression. Apart from beamforming, DoA estimation
becomes an inherent part of the overall architecture.  Typically, the
signal that is received using spatial sensors is considered a
sum-of-complex-exponentials along with additive white Gaussian
noise. This particular model appears in a variety of applications,
such as finite rate of innovation (FRI) sampling and
denoising~\cite{vetterli2002sampling}, target identification and
classification~\cite{sarkar1995using,sathe2022automatic}, and
estimation of graph-dynamics~\cite{venkitaraman2022annihilation}. With
this model, the problem reduces to estimating the ``frequency''
parameter from the observed vector signal, as we discuss in
Section~\ref{sec:model}. Depending on the context of the problem, the
notion of the ``frequency'' parameter changes. Extensive prior work
has been done to estimate this frequency, using the MVDR method and
other approaches. This includes subspace based methods for DoA
estimation such as MUSIC~\cite{1143830}, ESPRIT~\cite{roy1989esprit}
and their
variants~\cite{khan2008analysis,elbir2020deepmusic,hwang2008direction,gao2005generalized,steinwandt2017generalized},
filter based
methods~\cite{duan2005new,yan2006optimal,duan2007broadband}, and
machine learning based
methods~\cite{salvati2016use,alkhateeb2018deep}.
Several variants of the conventional MVDR~\cite{van2002optimum} exist
to enhance performance. One such method is robust 
MVDR~\cite{stoica2002robust,li2003robust}, that introduces an extra 
regularising parameter in the MVDR weight optimisation,
that makes MVDR robust against the element-misalignments. 
Nested-array MVDR~\cite{zheng2019robust} is
an interesting method that combines the concept of coarray with MVDR
to enhance the degrees-of-freedom of the conventional method. Recently, MVDR using
reduced dimension with subarrays~\cite{liu2022reduced} has been proposed, that
divides the whole array into multiple subarrays, and performs MVDR for each of
them in a serial fashion. One common aspect 
among all these prior works is that the estimation is
based on a finite impulse response (FIR) model.

It is well-known that infinite impulse response (IIR) filters provide
some advantages over FIR filters for several applications. Unlike FIR
filters, an IIR filter's transfer function has a numerator
that accounts for the zeros of the transfer function and a denominator
that accounts for poles of the transfer function. Due to the combined
effect of both poles and zeros, IIR filters address some key issues
with FIR filters~\cite{oppenheim2001discrete} such as:
\begin{itemize}
	\item Potentially offering a lower filter order for comparable performance.
	\item Offering a sharper roll-off for passband-stopband transitions.
\end{itemize}
While IIR filters offer these advantages, there are also some
trade-offs that are involved when using IIR filters. These include a
non-linear phase response, more complexity in optimizing
coefficients etc. To this end, we ensure that the designs presented
here are not affected by these shortcomings~\cite{antonion1993digital}.

Prior work related to IIR filter based beamforming
include~\cite{duan2005new,duan2007broadband,wen2013extending,karo2020source}. Most
prior works consider direct implementation of the IIR filter by
replacing the delay-and-sum FIR structure. Restructuring the
delay-and-sum to individual IIR filters is explained
in \cite{duan2005new,duan2007broadband} wherein it is proposed to
replace the delays of the delay-and-sum with tap-delay IIR filters,
and the filter coefficients are designed using the least-mean-square
method iteratively. Extending this work in \cite{wen2013extending},
the ``spatial'' delay
elements are estimated using the recursive-least-squares method,
which then can be used to estimate the spatial frequencies
present in the signal. These architectures are
largely an approximation, and use time-domain IIR filtering to achieve
required desired spatial frequency response. To get an analogous spatial IIR structure,
the concept of ``spatial feedback'' has to be implemented, as
discussed in~\cite{karo2020source}.  Here, the authors propose to
achieve this feedback by continuously retransmitting the beamformed
signal to achieve IIR like performance.

In our work, we (a) consider that the retransmission is performed 
using an array instead of a single
element, and (b) propose a method to incorporate the MVDR beamforming with
this feedback structure to further improve the performance.  This is
an effective way to exploit the larger number of antennas available on
MIMO wireless communication systems for more efficient DoA
estimation. We first develop an optimal retransmission strategy that
effectively utilizes the multitude of antennas by maximizing the
Fisher information for the radar transceiver, and use this to infer
the DoA. Through simulations, we first compare the performance
parameters of the proposed method with conventional FIR beamformer and
the single-element feedback method proposed in~\cite{karo2020source},
and show that our method outperforms both, achieving higher
directivity, narrower beamwidth and improved side-lobe suppression. We
then compare the DoA estimation using the proposed method with
MUSIC~\cite{1143830}, ESPRIT~\cite{roy1989esprit}, robust MVDR 
beamforming~\cite{stoica2002robust,li2003robust}, nested-array MVDR 
beamforming~\cite{zheng2019robust}, and reduced dimension MVDR beamforming~\cite{liu2022reduced}
methods, and show that our method
works better even in low SNR scenarios. The proposed method is able to
provide beamforming output, and can be used to estimate target directions,
simultaneously.

The rest of the paper is organised as follows: Section~\ref{sec:model}
discusses the system model in the context of array
signal processing using a uniform linear array
(ULA). Section~\ref{sec:feedbackbf} explains the proposed feedback
beamforming structure using an array. Section~\ref{sec:perform}
discusses the performance parameters, and Section~\ref{sec:fbmvdr}
discusses the use of MVDR with feedback beamforming for DoA
estimation. Section~\ref{sec:results} shows the comparison of
beam-pattern and its performance parameters, and performance of DoA estimation
method with prior methods. Finally, Section~\ref{sec:conclude}
concludes the discussion.

\section{System Model}
\label{sec:model}
Consider the narrowband DoA estimation problem, where the reflected 
signal from $L$ number of targets is captured using an $N$-element 
uniform linear array (ULA). At any time $n$, the received
signal vector can be written as
\begin{equation}
	\vec{r}[n] = \sum_{k=1}^{L} a_k[n] \vec{v}(\psi_k) + \vec{w}[n].
	\label{eq1:model}
\end{equation}
Here, $\vec{v}(\psi_k) = \left[
1~\e{-j\psi_k}~\dots~\e{-j(N-1)\psi_k}\right]$, $\psi_k=\frac{2\pi
d}{\lambda}\cos{\theta_k}$ is the spatial frequency, $\lambda$ is the
operating wavelength, $d$ is the inter-element spacing of ULA, and
$\vec{w}$ is complex additive white Gaussian noise with zero mean and
variance $\sigma^2$. The spatial frequency $\psi_k$ depends on the
angle of arrival $\theta_k$ of the returned signal, and is measured
from the array axis. The targets are assumed to be
stationary, and the reflected signals from each target are assumed to
be uncorrelated with each other~\cite{sandhu1985real}.

The model shown in \eqref{eq1:model} is closely related to the sum-of-complex-exponential model
that has been used in several other problems,
cf.~\cite{vetterli2002sampling,roy1989esprit,sarkar1995using,sathe2022automatic}. 
Therefore, the solution discussed here can be generalised to different problems,
when presented in this form.

We aim to estimate $\psi_k$ from $\vec{r}$. As mentioned in
Section~\ref{sec:intro}, several methods exist to solve such
systems. In this work, we take the approach of estimating the
parameters $\psi_k$ using the
MVDR~\cite{van2002optimum} beamforming method, and incorporate it
with spatial IIR feedback beamforming~\cite{karo2020source}. We
compare the performance parameters of the beam-pattern, viz. beamwidth, first
side-lobe level and directivity, with the feedback beamformer and
conventional FIR beamformers that have been used in past work. We then
use the MVDR IIR structure to estimate $\psi_k$, and show that our
approach outperforms the existing methods in terms of requiring a
lower SNR, and possessing a smaller beamwidth, thus resulting in finer
resolution and higher side-lobe suppression.

\section{Feedback Beamforming}
\label{sec:feedbackbf}
There are multiple approaches for constructing spatial IIR beamformers. We
enumerate them below.
\begin{itemize}
	\item Case 1: No feedback (FIR)
	\item Case 2: Feedback without retransmission
	\item Case 3: Feedback with retransmission using a single antenna
	\item Case 4: Feedback with retransmission using an antenna array
\end{itemize}
Case 1 is the conventional beamformer, wherein the beamformer response
can be modelled similar to that of a discrete-time FIR filter. Case 2
was introduced in~\cite{wen2013extending}, where the delay-like
elements in the feedback depend on an initial estimate of $\psi_k$,
and are iteratively refined over time using approaches such like
recursive least-squares (RLS). Cases 3 and 4 involve retransmission
based methods, which is the closest to a discrete-time IIR filter
based approach.  Case 3 was proposed in~\cite{karo2020source}, where
the captured signal is retransmitted using a single transmitting
antenna. The approach in~\cite{karo2020source} is specific to the case
of a single transmit antenna, and does not extend directly to the case
of multiple antennas. The case of multiple antennas is, however,
interesting from a practical point of view, since
several recent radar applications involve situations where
multiple antennas exist and can be exploited to enhance
performance. Therefore, we present Case 4, which is the multi-antenna retransmission
concept, wherein the retransmission is performed using an
array, as opposed to a single element.

Overall, the goal is to design the beamformer weights based on an appropriate
optimality criterion. One way to achieve this is to maximise the Fisher
information of the beamformer output. Given \eqref{eq1:model} as the
signal captured by the antenna array, the beamformer output is
\begin{equation}
	y[n] = \vecsym{\beta}^H \vec{r}[n]
	\label{eq1:feedbackbf}
\end{equation}
where $\vecsym{\beta} = [\beta_0 ~\ldots ~\beta_{N-1}]^T$ is the beamformer
weight vector. The Fisher information in the output is
\begin{equation}
	J(\psi) = \Re\left \{\frac{1}{2\pi \sigma^2} \int_{-\omega_s/2}^{\omega_s/2} \left| \frac{\del Y(\e{j\omega})}{\del \psi} \right|^2 \del \omega \right\}
	\label{eq2:feedbackbf}
\end{equation}
where $Y(\e{j\omega})$ is the discrete-time Fourier transform of $y[n]$,
$\omega_s$ is the bandwidth of interest, $\sigma^2$ is noise spectral density, and $\Re\{\cdot\}$
denotes the real part. Maximising $J(\psi)$ with respect to $\vecsym{\beta}$ 
yields optimum beamformer weights.
We will use this approach to obtain 
the beamformer weights of the proposed method as well.

\subsection{Feedback with retransmission using single antenna}
\label{subsec:singlefb}
\begin{figure}[!hbt]
	\centering
	\includegraphics[width=\imgwidth]{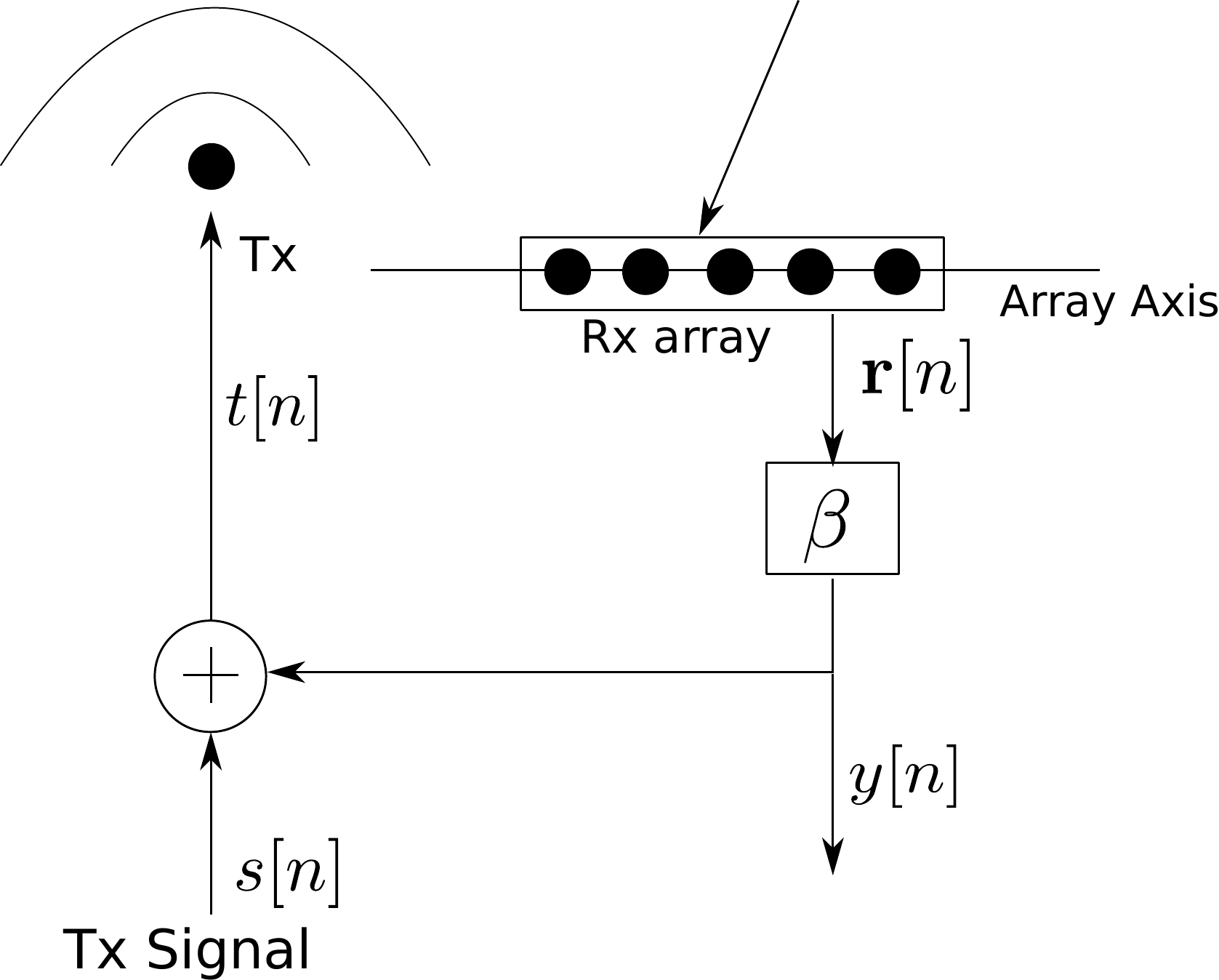}
	\caption{Feedback beamforming block diagram with single transmitting antenna.}
	\label{fig:singlefb_blk}
\end{figure}

The feedback beamforming as explained in \cite{karo2020source} is shown in \figurename{\ref{fig:singlefb_blk}}.
The signal in the feedback path is retransmitted using only a single isotropic
transmitting antenna. The diagram shown here is a simplified version
of what is shown in~\cite{karo2020source}, since the optimum feedback
coefficients obtained by maximising the Fisher
information matrix (FIM) is the same as the beamformer weights
$\vecsym{\beta}$. Therefore, combining both the paths result in the
structure shown in
\figurename{\ref{fig:singlefb_blk}}. The overall feedback beamformer response
is given as:
\begin{equation}
	H_{\text{single}}(\psi) = \frac{\vecsym{\beta}^H \vec{v}(\psi) }{1 - \vecsym{\beta}^H \vec{v}(\psi) }
	\label{eq1:singlefb}
\end{equation}
where $\vecsym{\beta} = \left[ \beta_0 ~\ldots ~\beta_{N-1}\right]^T$. 
The beamformer response in the case of a conventional FIR
beamformer is $H_{\text{FIR}}(\psi) = \vecsym{\beta}^H\vec{v}(\psi)$.
From \eqref{eq1:singlefb}, it can be seen that because of the feedback,
the beamformer response also has the denominator term when 
compared to $H_{\text{FIR}}(\psi)$. This
is explained in detail in ~\cite{karo2020source}. In the proposed work, we
focus on obtaining the angle ($\theta$) information only, and hence
we discard the parameter $\phi$ used in~\cite{karo2020source} that is 
associated with range estimation, and omit the estimation of signal
powers $P_k = \MB{E}(|a_k|^2)$.

\subsection{Retransmitting feedback with Array}
\label{subsec:arrayfb}
\begin{figure}[!hbt]
	\centering
	\includegraphics[width=\imgwidth]{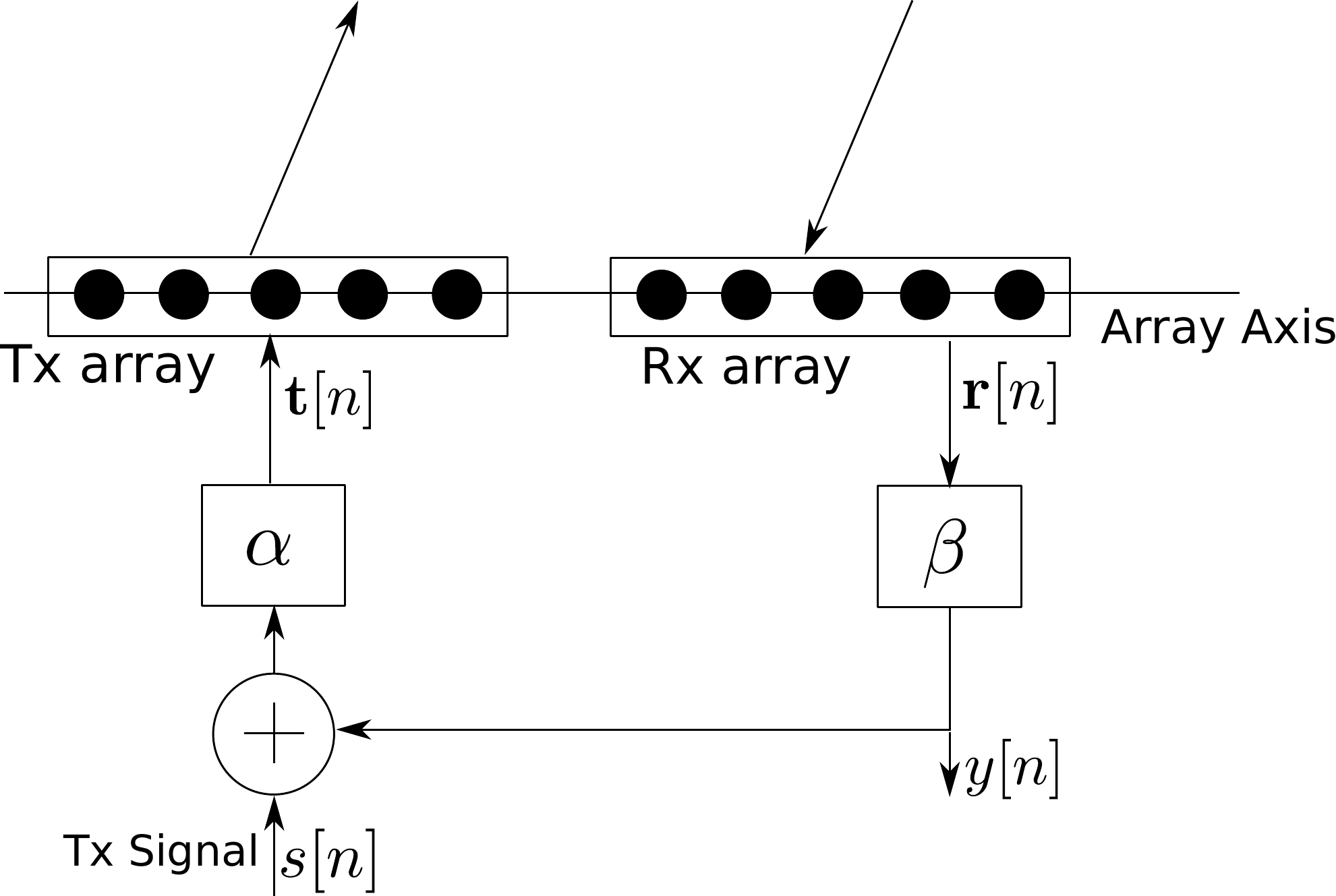}
	\caption{Feedback beamforming block diagram with transmitting antenna array.}
	\label{fig:arrayfb_blk}
\end{figure}

Extending the previous concept, we propose a retransmission method using an array instead
of a single element. The modified block diagram is shown in \figurename{\ref{fig:arrayfb_blk}}.

Unlike in the previous case, the transmit signal is directional,
since we are using an array instead of a single isotropic element. The
overall beamformer transfer function is given as:
\begin{equation}
	H_{\text{array}}(\psi) = \frac{\vecsym{\beta}^H \vec{v}(\psi) }{1 - \vecsym{\alpha}^H \vec{v}(\psi) \vecsym{\beta}^H \vec{v}(\psi) }.
	\label{eq1:arrayfb}
\end{equation}
Comparing \eqref{eq1:arrayfb} with \eqref{eq1:singlefb}, the
denominator possesses both $\vecsym{\alpha}$ and $\vecsym{\beta}$,
which contributes to higher directivity. In the case where there is
a single target at $\psi$, the optimal weights
$\vecsym{\alpha}$ and $\vecsym{\beta}$ are obtained by maximising the
FIM:
\begin{equation}
	\begin{split}
	\vecsym{\beta}(\psi) &= \frac{k\vec{v(\psi)}}{N},\\
	~\text{and}~\vecsym{\alpha}(\psi) &= \frac{\vec{v(\psi)}}{kN}, ~k \in \MB{C}-\{0\}.
	\end{split}
	\label{eq2:arrayfb}
\end{equation}
Both \eqref{eq1:arrayfb} and \eqref{eq2:arrayfb} are derived 
in Appendix~\ref{apdix:FIMderiv}.

\section{Performance parameters}
\label{sec:perform}
In this section, we derive the performance parameters of the proposed beamformer,
viz. Half-Power beamwidth (HPBW), First Side-Lobe Level (FSLL) and
directivity using the generalised beam-pattern expression,
\begin{equation}
	H_{\text{array}}(\psi) = \frac{g\vecsym{\beta}^H \vec{v}(\psi) }{1 - g\vecsym{\beta}^H \vec{v}(\psi) \vecsym{\alpha}^H \vec{v}(\psi) }
	\label{eq1:perform}
\end{equation}
where $g$ is the received signal gain. In the following
sections, we refer to $H_{\text{array}}(\cdot)$ as $H(\cdot)$ only. The beamformer weights 
$\vecsym{\alpha}$ and $\vecsym{\beta}$ are dependent on the
spatial frequency $\psi$ as shown in \eqref{eq2:arrayfb}.
Assuming that the target spatial frequency is $\psi_0$, the
beam-pattern can be obtained as a function of $\psi$:
\begin{equation}
	B(\psi) = |H(\psi)| = \left|\frac{\vecsym{\beta}^H(\psi) \vec{v}(\psi_0)}{1 - \vecsym{\beta}^H(\psi) \vec{v}(\psi_0) \vecsym{\alpha}^H(\psi) \vec{v}(\psi_0)}\right|.
	\label{eq2:perform}
\end{equation}
Using \eqref{eq2:arrayfb} in \eqref{eq2:perform},
\begin{equation}
	B(\psi) = \frac{\frac{\sin{\left(N(\psi-\psi_0)/2\right)}}{\sin{\left((\psi-\psi_0)/2\right)}}}{1 - \left(\frac{\sin{\left(N(\psi-\psi_0)/2\right)}}{\sin{\left((\psi-\psi_0)/2\right)}}\right)^2}
	\label{eq3:perform}
\end{equation}
which is a standard beam-pattern expression. If the feedback filter 
$\vecsym{\alpha}$ is removed, \eqref{eq2:perform} reduces to the conventional FIR
beamformer response $B_{\text{FIR}}(\psi) = |H_{\text{FIR}}(\psi)| = |\vecsym{\beta}^H\vec{v}(\psi)|$.

\subsection{Half-Power Beamwidth (HPBW)}
\label{subsec:hpbw}
Assuming the filter gain $\hat{g} \neq g$, we have
\begin{equation}
	\vecsym{\beta}(\psi) = \frac{\vec{v}(\psi)}{\hat{g}N}~\text{and}~\vecsym{\alpha}(\psi) = \frac{\vec{v}(\psi)}{kN}.
\end{equation}
Substituting values in \eqref{eq1:perform}, we obtain the maximum gain as
\begin{equation}
	H(\psi)\bigg|_{\text{max}} = \frac{\frac{g}{\hat{g}}}{1 - \frac{g}{k\hat{g}}} = \frac{r}{1 - \frac{r}{k}}
\end{equation}
where $k$ is tunable to ensure that the denominator can always be made
zero. Hence, ideally, the beamwidth remains zero regardless of the gain
$\hat{g}$, compared to the HPBW of single element feedback
~\cite{karo2020source} $\frac{1.4}{f(r)N}$, which is a
function of the gain mismatch $r = \frac{g}{\hat{g}}$.

\subsection{First Side-lobe Level}
\label{subsec:fsll}
The side-lobe (secondary) peaks of $H(\psi)$ are at
\begin{equation}
	\psi - \psi_0 = \frac{2m+1}{N}\pi,~\forall m \in \MB{Z}-\{0\}.
\end{equation}
Hence the first side-lobe is at $m=1$, which is $\Delta \psi
= \frac{3\pi}{N}$. For a large enough $N$, from \eqref{eq3:perform} we have
\begin{equation}
	\begin{split}
	\text{FSLL}(N) &= \left| r\frac{\frac{\sin{3\pi/2}}{\sin{3\pi/2N}}}{1 - \frac{r}{k}\left(\frac{\sin{3\pi/2}}{\sin{3\pi/2N}}\right)^2} \right|^2 \\ ~ & \approx \frac{9\pi^2k^2}{4 N^2} \left( 1+\frac{9\pi^2}{2 N^2} \right) \to \MC{O}(N^{-2})				
	\end{split}
	\label{eq:fsll_formula}
\end{equation}
which is independent of the gain mismatch $r$, and
\begin{equation}
	\lim_{N \to \infty} \text{FSLL}(N) \to 0.
\end{equation}

\subsection{Directivity}
\label{subsec:dir}
The maximum directivity $D$ is defined as
\begin{equation}
	D = \frac{H(\psi)|_{\text{max}}}{\frac{1}{2\pi}\int_{0}^{2\pi} H(\psi) \dif \psi} = \frac{\frac{2\pi r}{1 - \frac{r}{k}}}{\int_{0}^{2\pi} H(\psi) \dif \psi}.
\end{equation}
In this case, $H(\psi)|_{\text{max}}$ is tunable, hence directivity is a function of the tuning parameter $k$.

\section{Array feedback beamforming for DoA estimation using MVDR}
\label{sec:fbmvdr}
\begin{figure}[!hbt]
	\centering
	\includegraphics[width=\imgwidth]{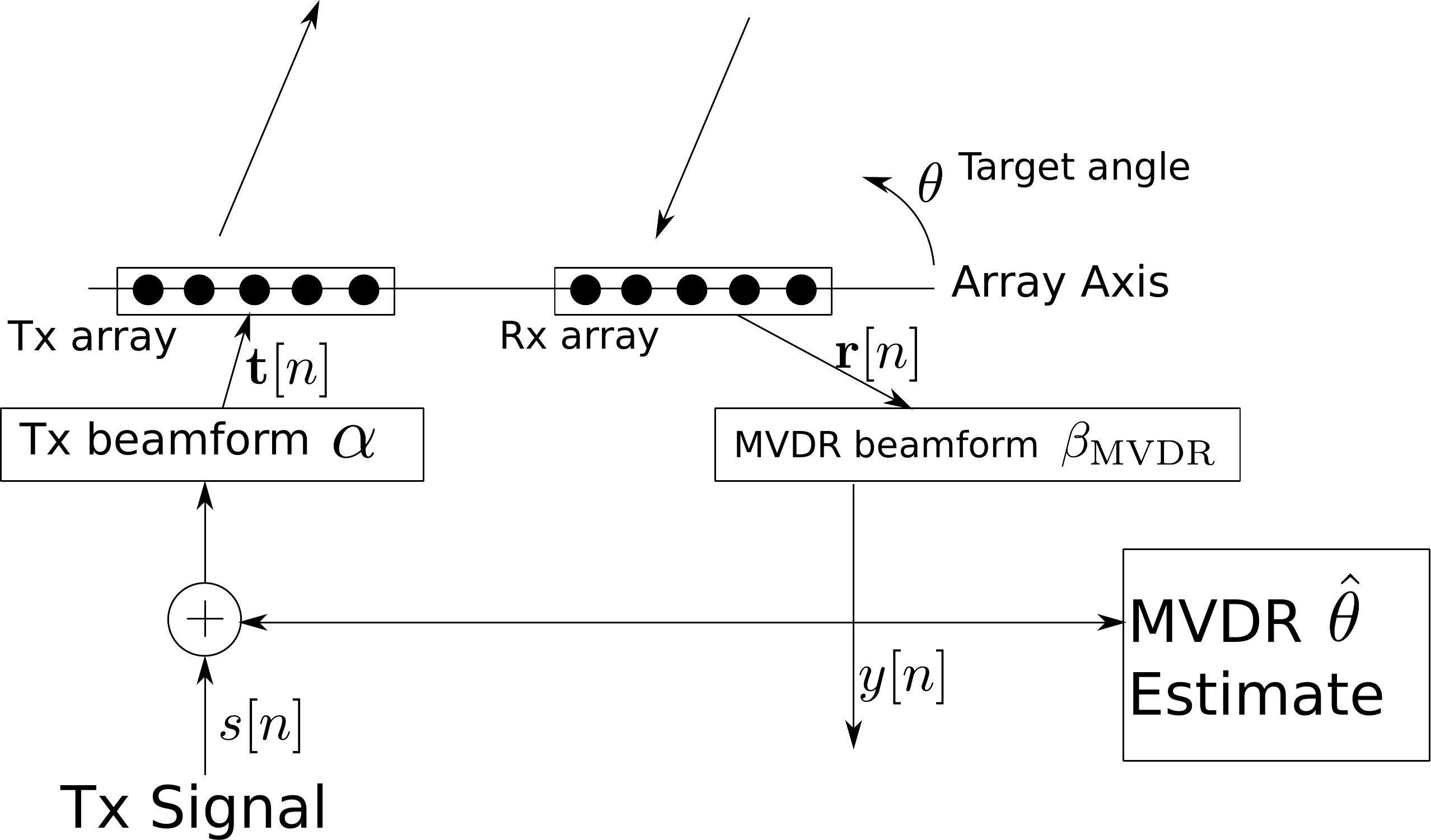}
	\caption{Block diagram of MVDR DoA estimation using feedback beamforming.}
	\label{fig:fbmvdr_block}
\end{figure}

We now present the proposed approach to estimate the $\psi$ parameters using
MVDR with feedback beamforming for the model shown
in \eqref{eq1:model}. \figurename{\ref{fig:fbmvdr_block}} shows the
block diagram depicting how MVDR can be used with a feedback
beamforming structure.

As shown in \figurename{\ref{fig:fbmvdr_block}}, the signal is first captured using
a ULA, and the received signal $\vec{r}$ as in \eqref{eq1:model} is used to compute
the MVDR coefficients. Given the autocorrelation matrix 
$\vec{R}_{\vec{rr}} = \frac{1}{N_{\text{samples}}}\sum_{n=1}^{N_{\text{samples}}}\vec{r}[n]\vec{r}^H[n]$ and a
linear constraint $\vecsym{\beta}^H\vec{c} = 1$, the MVDR coefficients can be computed as:
\begin{equation}
	\vecsym{\beta}_{\text{MVDR}} = \frac{\vec{R}^{-1}_{\vec{rr}}\vec{c}}{\vec{c}^H \vec{R}^{-1}_{\vec{rr}} \vec{c}}
	\label{eq1:fbmvdr}
\end{equation}
where $N_{\text{samples}}$ are the available number of time-samples.

Next, the feedback (retransmission) coefficients $\vecsym{\alpha}$ are
inferred from $\vecsym{\beta}$, and over multiple such
retransmissions, the coefficients get stabilised.  At this point,
$\psi$ parameters are estimated using $\vecsym{\alpha}$ and
$\vecsym{\beta}$.  Depending on the approach taken to infer
$\vecsym{\alpha}$ from $\vecsym{\beta}$, there are two
possible methods to estimate $\vecsym{\alpha}$, as shown in Algorithm~\ref{alg:method1} and 
Algorithm~\ref{alg:method2}.

The classical MVDR method was primarily designed to mitigate/suppress
interference, but here we utilise the same functionality to find the
target directions instead. Typically, the MVDR weights are computed in the
absence of the targets and in the presence of interference and jammers, so
that the algorithm steers the beamformer nulls in the directions where
the interferers and jammers are present while maintaining
distortionless response in a desired direction. However, in the
presence of targets, the MVDR method computes the beamformer weights such
that nulls are placed along the target directions. The constraint vector in \eqref{eq1:fbmvdr}
is taken as $\vec{c} = [1 ~\ldots ~\e{-j(N-1)\psi}]^T$ that
allows an undistorted response from the desired direction. This constraint
is used for Algorithm~\ref{alg:method1}.
\begin{algorithm}[!hbt]
	\caption{}
	\begin{algorithmic}[1]
	\State{Start with $\vecsym{\beta}$ as in \eqref{eq1:fbmvdr}.}
	\State{Formulate $\vecsym{\alpha} = \vec{c} = [1 ~\e{-j\psi} ~\dots ~\e{-j(N-1)\psi}]^T$}
	\State{Sweep $\psi$ in the visible region (typically $[-\pi,\pi)$). If a target angle
	coincides with the sweep angle, the response at that angle peaks while suppressing
	contribution of all the other targets.}
	\State{Transmit vector = $\vecsym{\alpha}(\psi)$} 	
	\end{algorithmic}
	\label{alg:method1}
\end{algorithm}

The same constraint vector $\vec{c}$ can be
changed to any other linear constraint vector, as long as the algorithm
produces computable beamformer weights. A simple choice is
$\vec{c} = [1 ~0 ~\ldots ~0]^T$, that ensures
the first element of $\vecsym{\beta}$, that is $\beta_0$,
is always $1$. This constraint is used for Algorithm~\ref{alg:method2}.
\begin{algorithm}[!hbt]
	\caption{}
	\begin{algorithmic}[1]
	\State{Start with $\vecsym{\beta}$ as in \eqref{eq1:fbmvdr} using $\vec{c} = [1 ~\ldots ~0]^T$. 
	The nulls of $\vecsym{\beta}$ appear along the target directions.}
	\State{Formulate $\vecsym{\alpha}$ as impulse response of $1/\vecsym{\beta}(\psi)$}
	\State{The nulls in $\vecsym{\beta}(\psi)$ result as peaks in $\vecsym{\alpha}(\psi)$}
	\State{Transmit vector = $\vecsym{\alpha}(\psi)$}
	\end{algorithmic}
\label{alg:method2}
\end{algorithm}

Both methods yield iteratively better estimates, even if the initial estimates 
are inaccurate. We discuss and compare the performance of both methods in the next section.

\section{Results and Discussion}
\label{sec:results}
In this section, via simulations, we first show the beam-pattern and
FSLL, as discussed in Section~\ref{sec:perform} for the proposed beamforming
method. We then show the performance of the DoA estimation using the proposed 
method and compare it with previous methods, viz. 
MUSIC~\cite{1143830}, ESPRIT~\cite{roy1989esprit}, Robust Capon beamforming~\cite{stoica2002robust,li2003robust},
nested array beamforming~\cite{zheng2019robust}, and reduced dimension beamforming~\cite{liu2022reduced}.

\subsection{Beam-pattern and performance parameters}
We consider a 3-element ULA, with inter-element spacing $\lambda/2$, and assume
a target present at $\theta = \pi/3$. We plot the beamformer response for
all angles $\theta \in [0,\pi)$ and compare the beam-pattern parameters of the proposed
method with that of existing methods.
\begin{figure}[!hbt]
	\centering
	\includegraphics[width=\imgwidth]{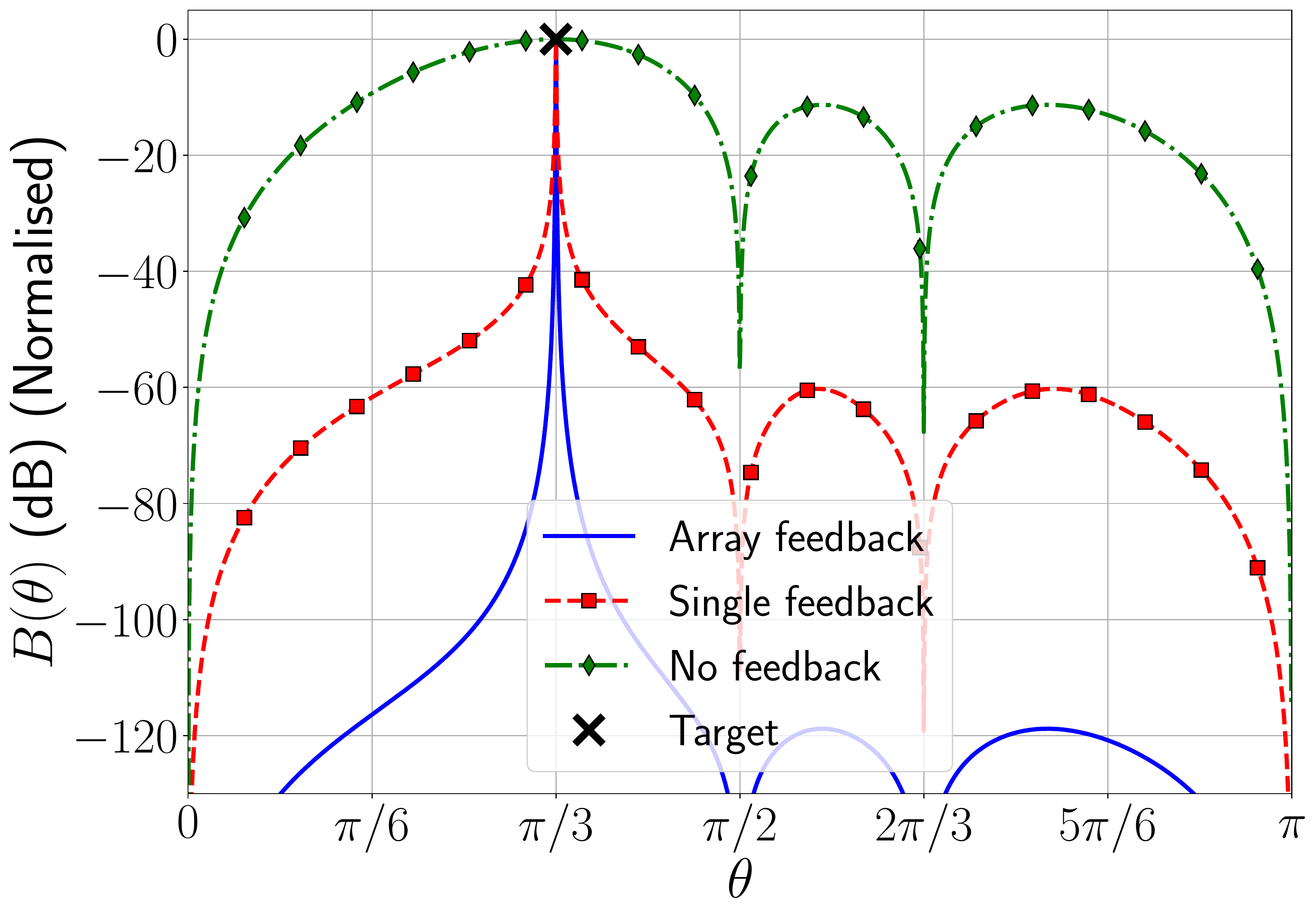}
	\caption{Comparing beam-pattern of array feedback beamformer with single feedback and conventional FIR beamformers}
	\label{fig:beamcompare}
\end{figure}

In \figurename{~\ref{fig:beamcompare}}, we compare the 
beam-pattern of the proposed feedback beamforming with existing 
feedback beamforming methods~\cite{karo2020source} and the conventional FIR method.
The HPBW of the FIR beamformer is limited by the number 
of elements, and that of the feedback beamformer is a
function of gain mismatch. Compared to the single element feedback case,
the HPBW of the proposed approach is 50\% less, which is consistent
with the reduction predicted in theory.
The side-lobe level of the feedback beamformer is 50~\db\
below conventional FIR beamformer, while that for the proposed method,
it is even lower, at $110$~\db. The proposed method outperforms 
both the conventional FIR beamformer and feedback beamformer~\cite{karo2020source} 
in terms of side-lobe suppression and beamwidth.

\begin{figure}[!hbt]
	\centering
	\includegraphics[width=\imgwidth]{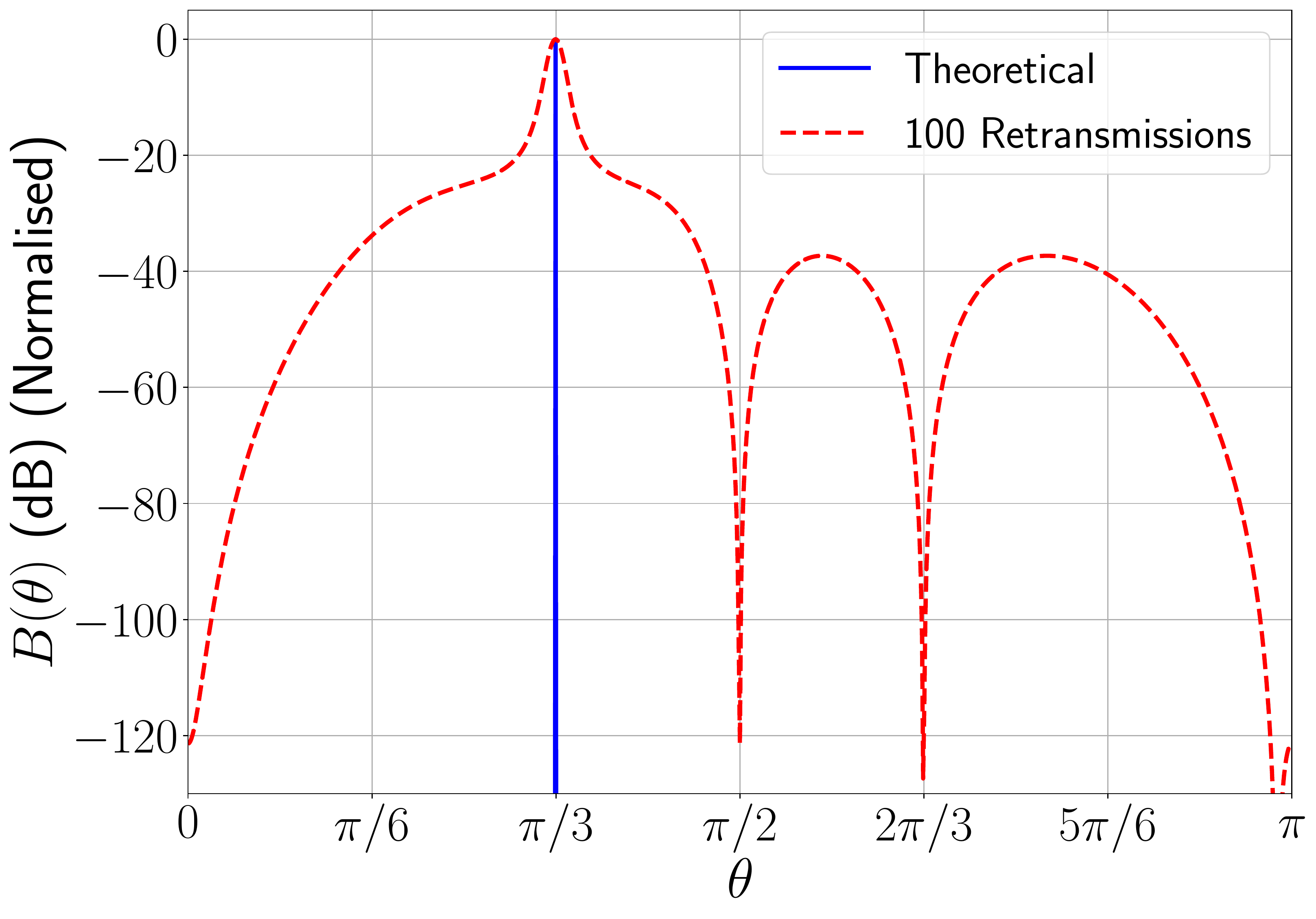}
	\caption{Beam-pattern after finite (100) retransmission time-stamps}
	\label{fig:beamfinite}
\end{figure}
\figurename{~\ref{fig:beamfinite}} shows the comparison of an ideal
feedback beamformer with that of a realisable one. In practice, the
IIR-like performance can be achieved only approximately, because of
the finite number of snapshots that are available. Even then, the
proposed method achieves an HPBW that is 90\% lower than that obtained
using FIR beamformers.

\begin{figure}[!hbt]
	\centering
	\includegraphics[width=\imgwidth]{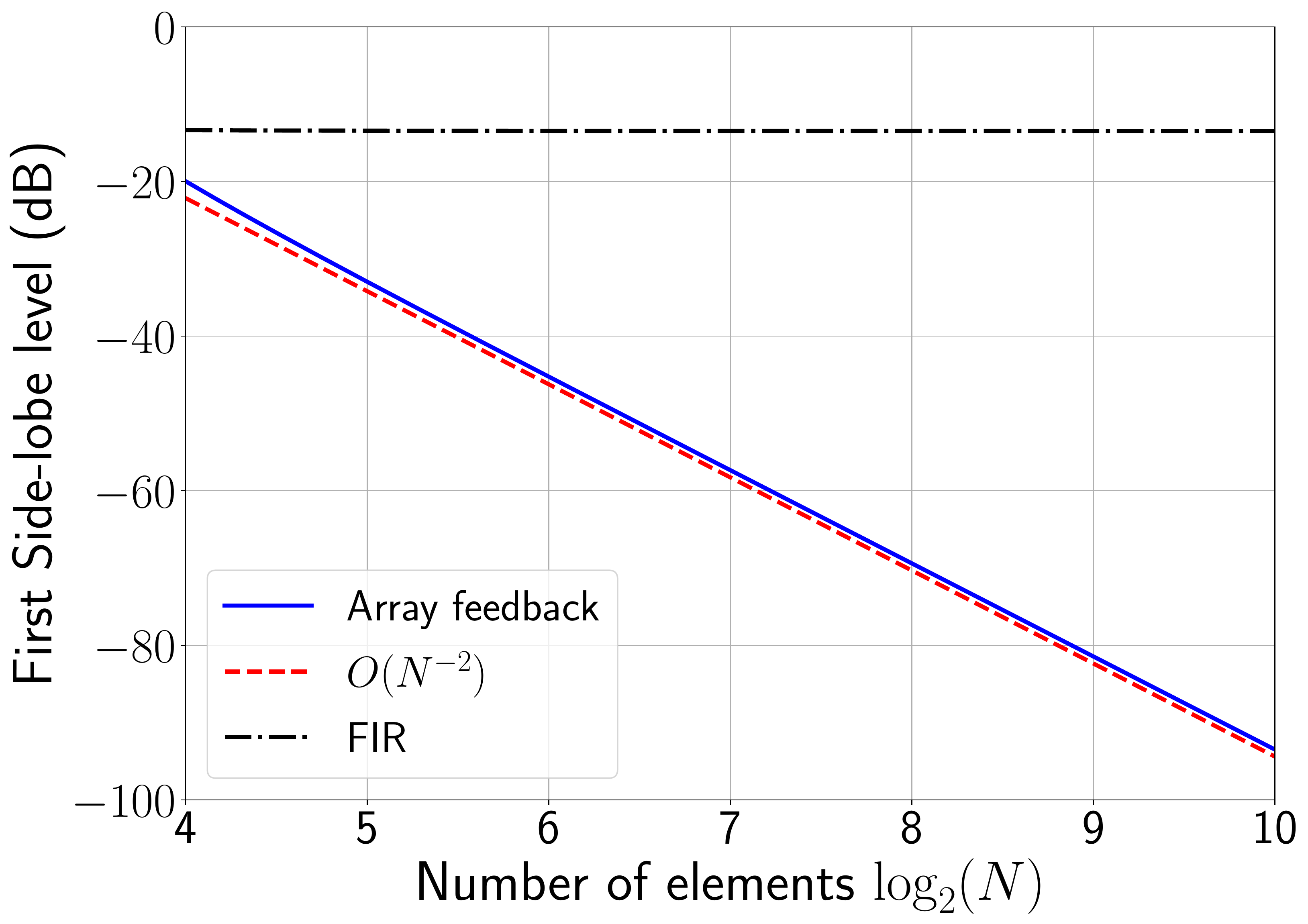}
	\caption{First Side-lobe level trend with number of array elements.}
	\label{fig:fsll}
\end{figure}
\figurename{~\ref{fig:fsll}} shows the absolute value of FSLL for 
different numbers of array elements. According to 
\eqref{eq:fsll_formula}, the FSLL decreases as 
$\MC{O}(N^{-2})$, and is independent of the gain mismatch $r$, as opposed
to the FIR beamformer, where the FSLL remains constant at 
$\sim -13.5$~\db. Even with as few as 16 elements, the level
difference is $7$~\db, and it decreases fast with increasing number
of elements. At 1024 elements, the difference is around $100$~\db.

\subsection{Direction Estimation}
As discussed in Section~\ref{sec:fbmvdr}, we can use the MVDR with 
feedback beamforming to estimate $\psi$ from
$\vec{r}$ in \eqref{eq1:model}. We use Algorithm~\ref{alg:method1}
and Algorithm~\ref{alg:method2} shown in
Section~\ref{sec:fbmvdr} and compare the resultant estimation error
with MUSIC and ESPRIT for various SNR levels. We
also consider inter-element spacing in the ULA $d
= \lambda/2$, hence $\psi = \pi\cos{\theta}$.
We take an 8-element ULA, and assume 4 targets at distinct angles
$\theta_1, \theta_2, \theta_3, \theta_4$. The signals reflected from
these targets are considered uncorrelated with each other. The received
signal at the array is assumed to follow the model in \eqref{eq1:model}.
\begin{figure}[!hbt]
	\centering
	\subfloat[Algorithm~\ref{alg:method1}]{\includegraphics[width=\twidth]{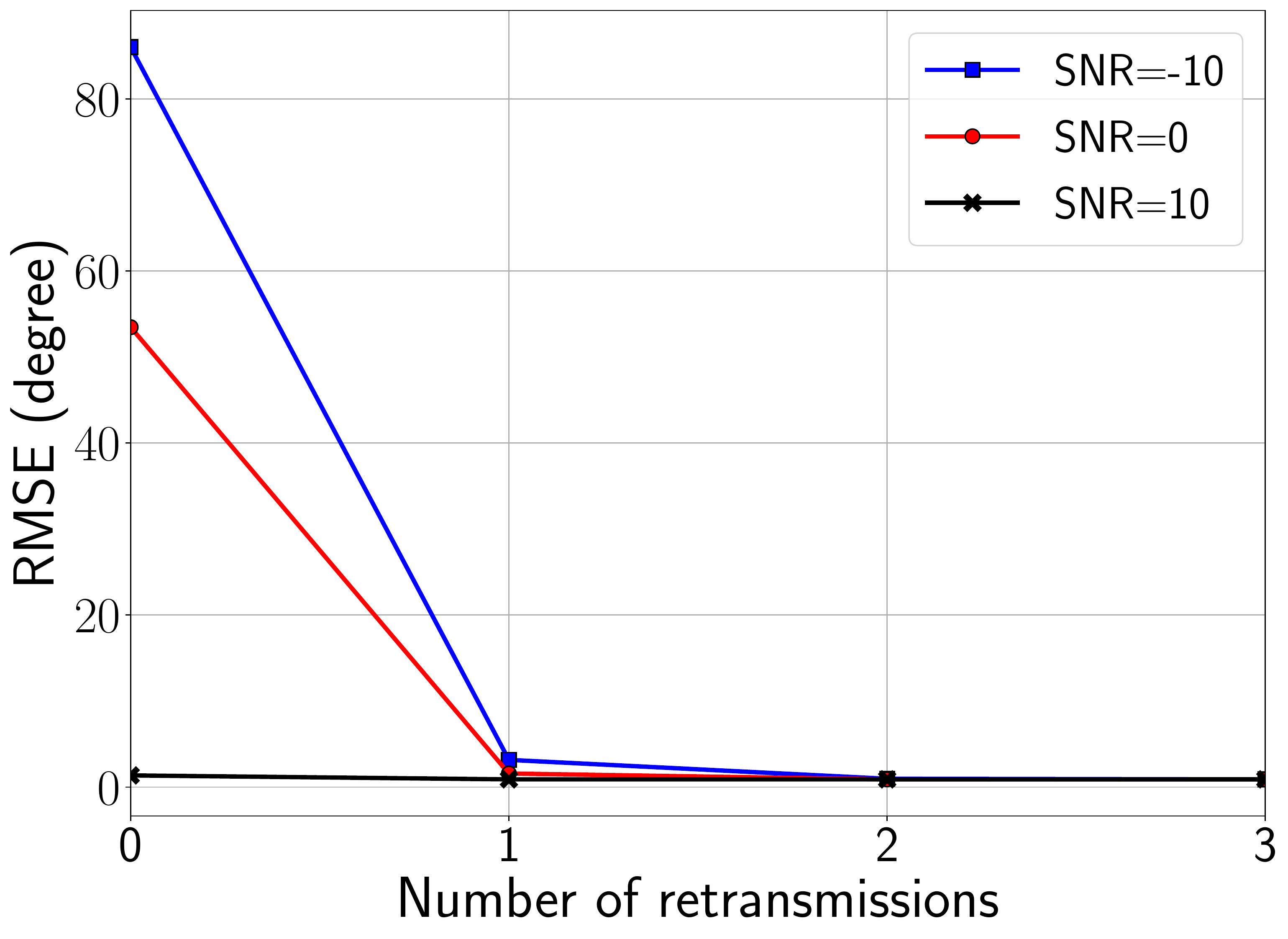}\label{sfig:fbmvdr_err2}}
	\subfloat[Algorithm~\ref{alg:method2}]{\includegraphics[width=\twidth]{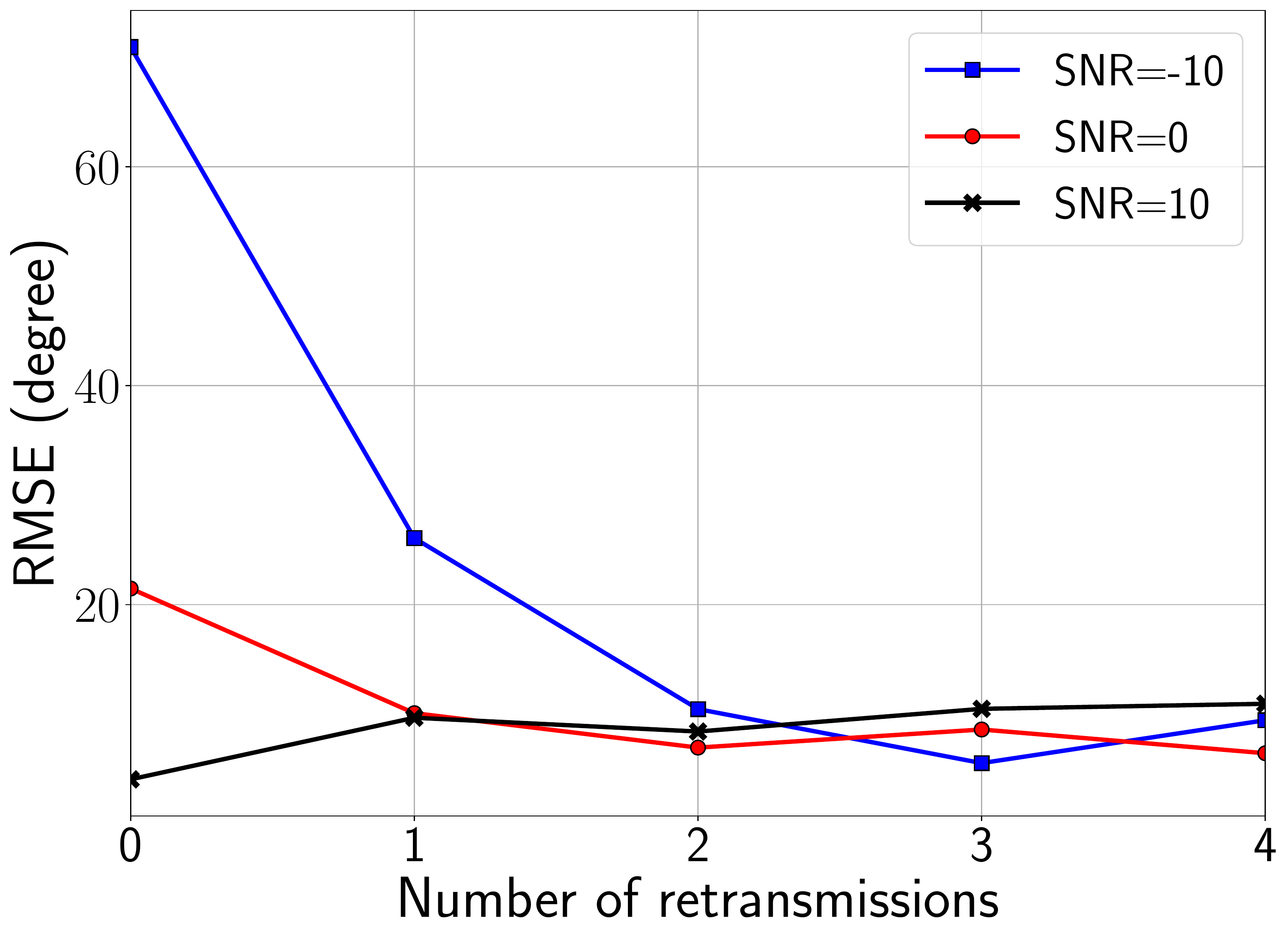}\label{sfig:fbmvdr_err1}}
	\caption{RMS error of MVDR DoA estimation using feedback beamforming with respect to number of retransmissions for SNR
	-10,~0,~and 10 \db}
	\label{fig:fbmvdr_err}
\end{figure}

\figurename{~\ref{fig:fbmvdr_err}} shows the root-mean-squared error (RMSE) for 
three different levels of SNR. Since our goal is to target the low
SNR region, we consider SNR = -10~\db, 0~\db,
and 10~\db. \figurename{~\ref{sfig:fbmvdr_err1}}
and \figurename{~\ref{sfig:fbmvdr_err2}} show the performance for
Algorithm~\ref{alg:method1} and Algorithm~\ref{alg:method2} discussed in Section~\ref{sec:fbmvdr},
respectively. From \figurename{~\ref{fig:fbmvdr_err}} it can be seen that, for no 
retransmission, the estimation error is large, since feedback is absent.
As the number of retransmissions increases, the error reduces rapidly for
both methods. Comparing \figurename{~\ref{sfig:fbmvdr_err1}} and
\figurename{~\ref{sfig:fbmvdr_err2}}, it is evident that Algorithm~\ref{alg:method1}
yields better estimates than Algorithm~\ref{alg:method2}, since the error reduces faster
for Algorithm~\ref{alg:method1}.
\begin{figure}[!hbt]
	\centering
	\subfloat[SNR=40~\db]{\includegraphics[width=\twidth]{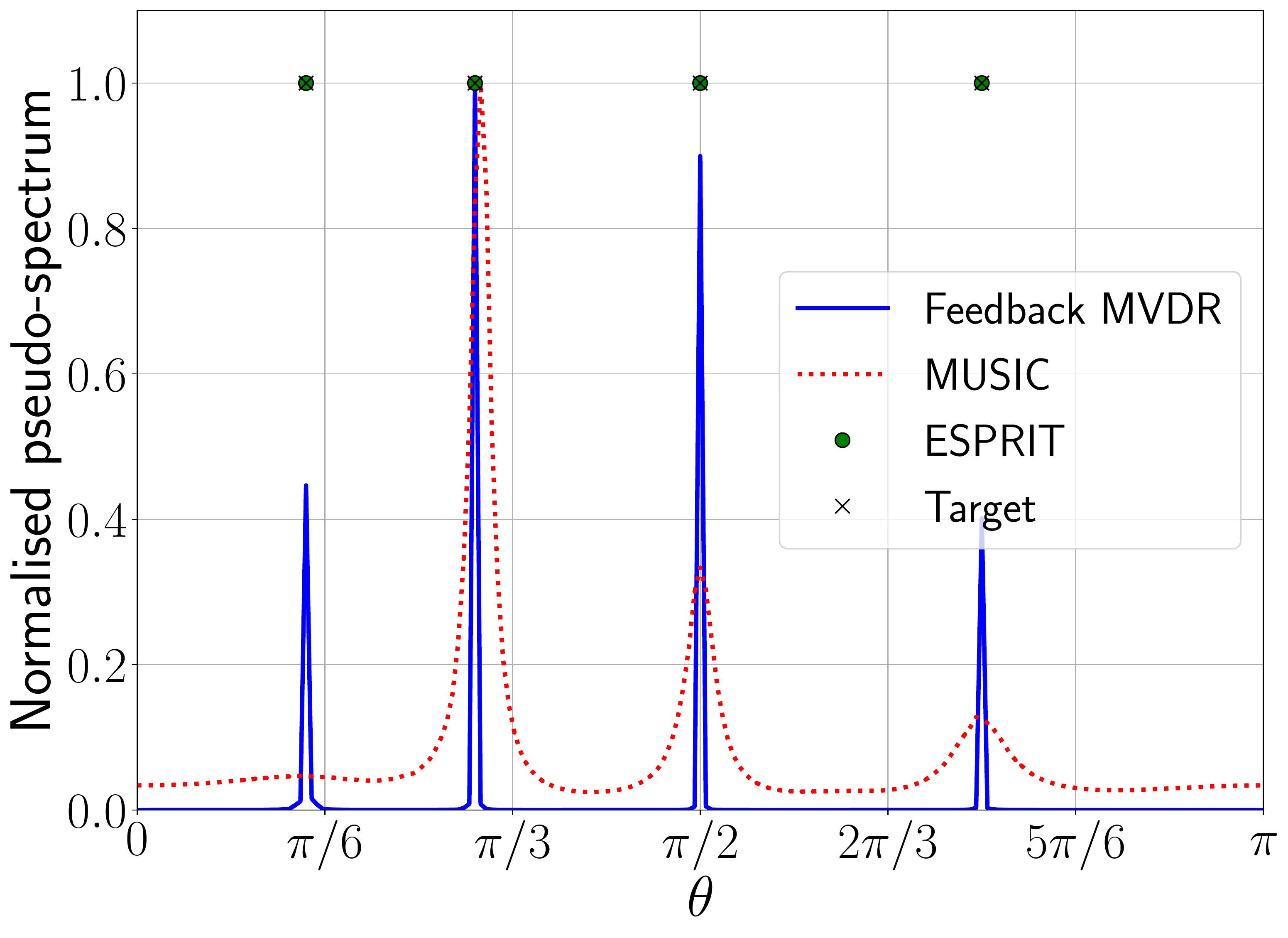}\label{sfig:fbmvdr_compare1}}
	\subfloat[SNR=0~\db]{\includegraphics[width=\twidth]{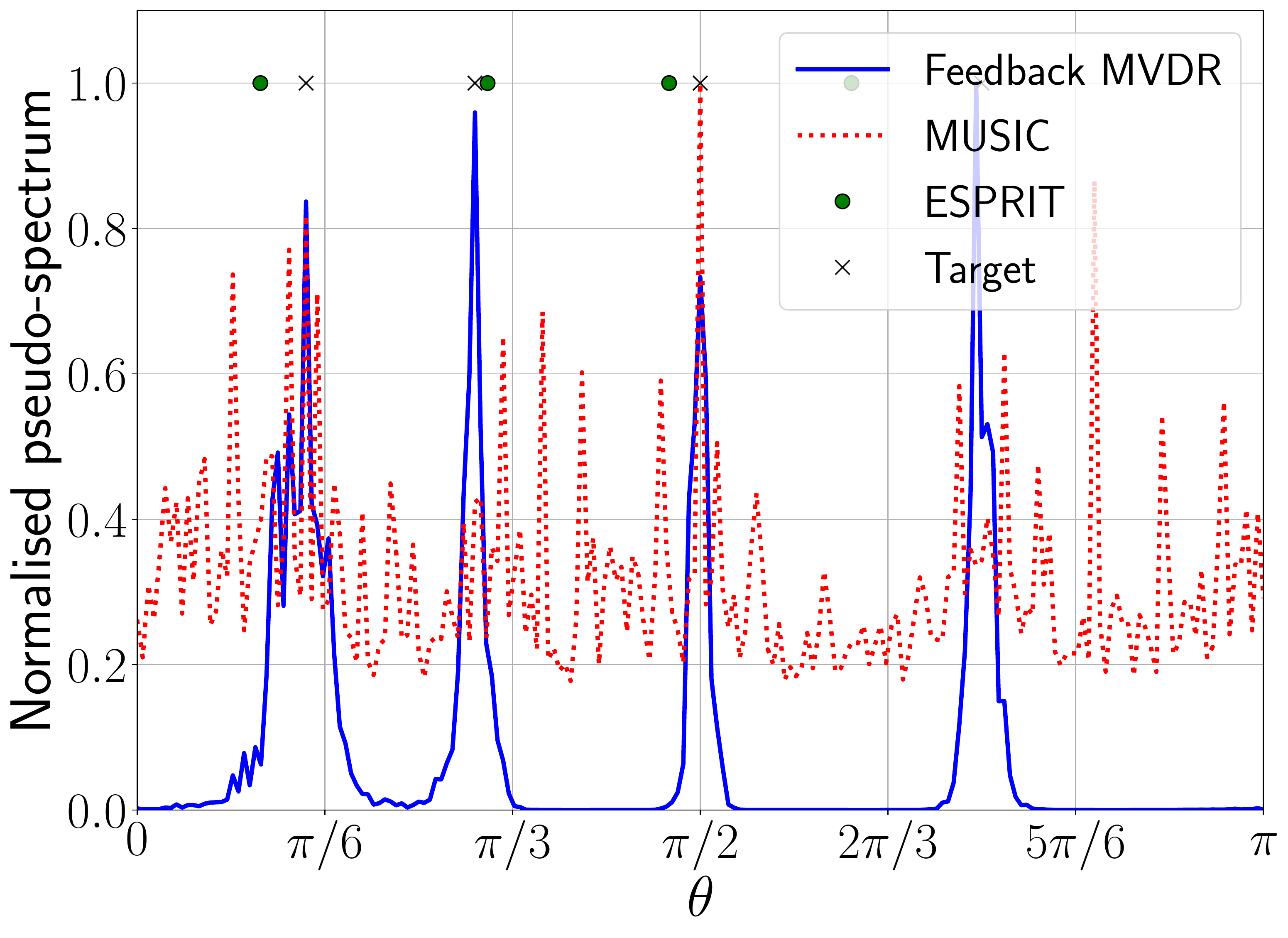}\label{sfig:fbmvdr_compare2}}
	\caption{Comparison of feedback MVDR with 2 retransmissions with MUSIC and ESPRIT}
	\label{fig:fbmvdr_compare}
\end{figure}

\figurename{\ref{fig:fbmvdr_compare}} shows the angle estimation
performance compared to MUSIC and ESPRIT.  For this, we use Algorithm~\ref{alg:method1}.
The array steering direction $\theta$ is varied from $0$ to $\pi$,
and the transmit weight vector $\vecsym{\alpha}$ varies
accordingly with the constraint $\vec{c}$. 
The beamformer response is computed as $y_{\psi}$, and
the pseudo-spectrum is computed as
$P_{\text{out}}(\psi) = \frac{1}{N_{\text{samples}}}\sum_{n}y_{\psi}[n]y_{\psi}^*[n],~\forall \psi$. 
\figurename{\ref{fig:fbmvdr_compare}} shows the normalised 
pseudo-spectrum as a function of $\theta$.

It can be seen that, with as low as just two retransmissions, the
feedback beamformer is able to achieve angle estimates at par with
that of ESPRIT at high SNR. However, for both MUSIC and ESPRIT, the number
of targets $L$ is known, which is not a constraint for the
proposed feedback MVDR method. Even at $0$~\db\ SNR, the proposed method shows
better estimates than both MUSIC and ESPRIT.

\begin{figure}[!hbt]
	\centering
	\includegraphics[width=\imgwidth]{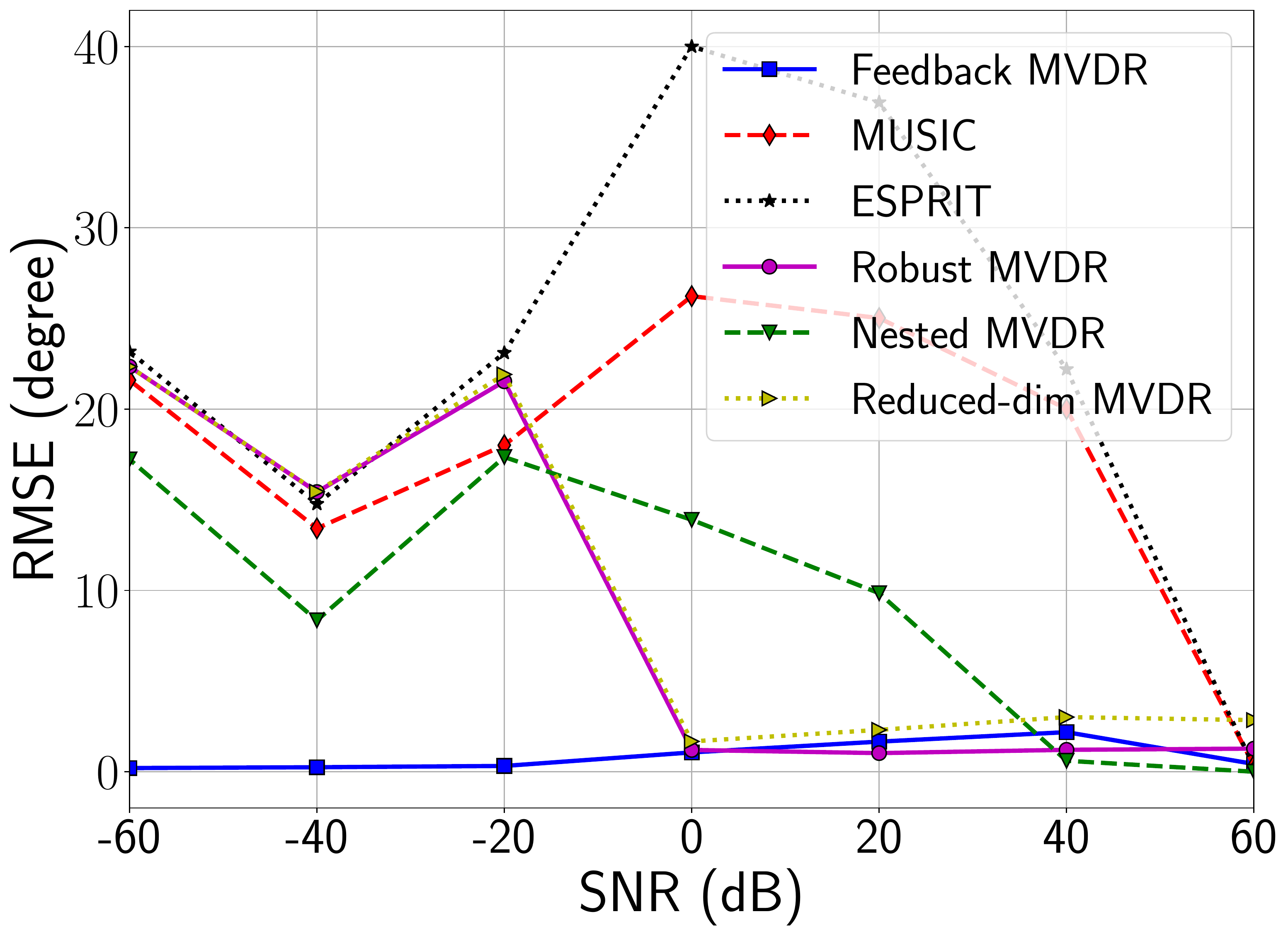}
	\caption{Comparison of RMSE with respect to SNR of feedback MVDR with
	prior methods.}
	\label{fig:doa_snr_compare}
\end{figure}

Finally, in \figurename{\ref{fig:doa_snr_compare}} we compare the
DoA estimation error of the proposed method
with previous methods for different SNR values in the following manner.
We consider $N=8$-element ULA with $d=\lambda/2$ inter-element spacing.
We assume two targets at $\theta_1$ and $\theta_2$, and the return signal
is modelled as in \eqref{eq1:model}. For MUSIC~\cite{1143830} and 
ESPRIT~\cite{roy1989esprit}, we consider this ULA and find the estimates
$\hat{\theta}_1$ and $\hat{\theta}_2$. For robust MVDR~\cite{stoica2002robust,li2003robust},
we assume the regularising parameter $\lambda_r=0.05$, which is suitably chosen
within the bounds~\cite{li2003robust}, and then $\hat{\theta}_1$ and $\hat{\theta}_2$
are estimated. For Nested array~\cite{zheng2019robust} setup, we consider
level-1 subarray of $N_1 = \floor{N/2} = 4$ elements with inter-element
spacing $d_1 = d = \lambda/2$ and level-2 subarray of $N_2 = \ceil{N/2} = 4$ 
elements with inter-element spacing $d_2 = (N_1+1) d = 5\lambda/2$, and then
perform the necessary rearrangement of the elements of the autocorrelation matrix
$\vec{R}_{\vec{rr}}$ to estimate $\hat{\theta}_1$ and $\hat{\theta}_2$. For
reduced dimension MVDR~\cite{liu2022reduced}, we consider $4$ subarrays of
$L=2$ elements each from the ULA, and use Algorithm~1 (SAMVDR) from
\cite{liu2022reduced} to estimate $\hat{\theta}_1$ and $\hat{\theta}_2$.
We then find the root-mean-squared error as 
$\sqrt{\sum_{k=1}^{2} (\theta_k - \hat{\theta}_k)}$ and take an average
over 100 Monte-Carlo runs.

From \figurename{\ref{fig:doa_snr_compare}} it is evident that feedback MVDR 
yields better error performance than all considered methods. Robust MVDR yields
better estimation error after $0$~\db\ SNR, while all other methods perform
better only after $40$~\db\ SNR. While these methods
primarily work better at higher SNR levels,
our method works even at the low SNR range of $-60$~\db\ to $-10$~\db. The 
RMSE of the proposed method is $20^{\circ}$ less than all methods in the low SNR region.

\section{Conclusion}
\label{sec:conclude}
We propose a novel approach to the problem of spatial filtering, beamforming
and DoA estimation using the feedback beamforming method with a retransmitting array.
We apply this approach to a ULA to derive the beam-pattern performance parameters.
Through extensive simulations, we show that our method outperforms the
conventional FIR beamformer and previously presented IIR beamforming structures in terms
of better side-lobe suppression, higher directivity, and narrower beamwidth. Then, we propose
a method to incorporate MVDR into this feedback structure and show that the combined
architecture is able to provide beamforming output, and can be used to estimate
the target directions. The proposed method outperforms  conventional DoA estimations methods
MUSIC and ESPRIT, and other MVDR variants, in terms of achieving better accuracy and better 
target separation. The proposed method is arguably better in the low-SNR regime, and yields 
$20^{\circ}$ lower estimation error, and hence is useful for applications with stringent 
SNR constraints.

\appendix
\label{apdix:FIMderiv}
Here we will derive the beamformer weights $\vecsym{\alpha}$ and $\vecsym{\beta}$
by maximising the Fisher information matrix. Consider the signal
$s[n]$ that is transmitted
using the $N_t$-element transmitting array, along with the feedback component.
The transmitted signal from the $k$th element of the array at time $n$ is given as
\begin{equation*}
	t_k[n] = \delta_k s[n] + \alpha_k^* y[n - \tau_k], \quad\delta_k = \begin{cases}
	                                                   	1&:~k=0 \\
	                                                   	0&:~\text{otherwise}
	                                                   \end{cases}
\end{equation*}
where $k=0,\ldots,N_t-1$, and $(\cdot)^*$ denotes complex conjugate. Similarly, 
the signal received from a single target located at $\psi$ at the $l$th element of the 
$N_r$-element receiving array is given as
\begin{align*}
	r_l[n] &= \sum_k t_k[n-\tau_l - \tau_R] \\
	 ~ &= s[n- \tau_l - \tau_R] + \sum_k \alpha_k^* y[n - \tau_k-\tau_l - \tau_R]
\end{align*}
where $k=0,\ldots,N_t-1$ and $l=0,\ldots,N_r -1$. $\tau_l$ and
$\tau_k$ are the delays due to the array
geometry, and $\tau_R$ is the delay due to the target range. Converting $r_l$ to frequency domain,
\begin{align*}
	R_l(\e{j\omega}) = \left( S(\e{j\omega}) + \sum_k \alpha_k^* \e{-j\omega\tau_k}Y(\e{j\omega})\right)\e{-j\omega (\tau_l+\tau_R)}
\end{align*}
where $R_l(\e{j\omega})$, $S(\e{j\omega})$, and $Y(\e{j\omega})$ are
the discrete-time Fourier transforms of
$r_l[n]$, $s[n]$ and $y[n]$, respectively.

The beamformer output is given as
\begin{align*}
	Y(\e{j\omega}) &= \sum_l \beta_l^* R_l(\e{j\omega}) \\
				~&= \left( S(\e{j\omega}) + \sum_k \alpha_k^* \e{-j\omega\tau_k}Y(\e{j\omega})\right) \sum_l \beta_l^* \e{-j\omega \tau_l}\e{-j\omega\tau_R}.
\end{align*}

Hence, the transfer function at $\omega$ is given as
\begin{align*}
	H(\e{j\omega}) = \frac{Y(\e{j\omega})}{S(\e{j\omega})} = \frac{\sum_l \beta_l^* \e{-j\omega\tau_l} \e{-j\omega\tau_R}}{1 - \sum_l \beta_l^* \e{-j\omega\tau_l} \sum_k \alpha_k^* \e{-j\omega\tau_k}\e{-j\omega\tau_R}}.
\end{align*}

Considering the narrowband signal model, for ULA, $\tau_l = l \tau$ and 
$\tau_k = k \tau$ for $k,l = 0,1,\dots, N-1$
assuming $N_t = N_r = N$. Taking 
$\omega\tau = \frac{2\pi}{\lambda}d\cos{\theta} = \psi$, and single frequency $\omega_0$
\begin{equation*}
	H(\psi,\phi) = \frac{\vecsym{\beta}^H \vec{v}(\psi) \e{-j\phi}}{1 - \vecsym{\beta}^H \vec{v}(\psi) \vecsym{\alpha}^H \vec{v}(\psi) \e{-j\phi}}
\end{equation*}
where $\vec{v}(\psi) = [1 ~\e{-j\psi} ~\e{-2j\psi} ~\dots ~\e{-j(N-1)\psi}]^T$
and $\phi = \omega\tau_R$.
This expression is shown in \eqref{eq1:arrayfb}.

We formulate the Fisher Information Matrix as below:
\begin{equation*}
	\vec{J} = \begin{bmatrix} J_{\psi\psi} & J_{\psi\phi} \\ J_{\phi\psi} & J_{\phi\phi} \end{bmatrix}
\end{equation*}
where $J_{\psi\phi} = J_{\phi\psi}^*$. Each term $J_{pq}$ can be computed as
\begin{equation*}
	J_{pq} = \Re\left\{\frac{1}{2\pi \sigma^2} \int_{-\frac{\omega_s}{2}}^{\frac{\omega_s}{2}} \left( \frac{\del Y(\e{j\omega})}{\del p} \right)^* \left( \frac{\del Y(\e{j\omega})}{\del q} \right) \dif \omega \right\}
\end{equation*}
where $\Re\{\cdot\}$ denotes real part, $p,q \in \{\phi,\psi \}$, $\omega_s$ is the bandwidth, 
and $\sigma^2$ is the noise spectral density. Hence,
\begin{footnotesize}
\begin{align*}
	\frac{\del Y(\e{j\omega})}{\del \psi} &= \frac{\vecsym{\beta}^H (\vec{I} + \vec{v}(\psi)\vecsym{\beta}^H\vec{v}(\psi)\vecsym{\alpha}^H \e{-j\phi}) \vec{A}\vec{v}(\psi)\e{-j\phi}}{(1 - \vecsym{\beta}^H \vec{v}(\psi) \vecsym{\alpha}^H \vec{v}(\psi)\e{-j\phi})^2}S(\e{j\omega}) \\
	\frac{\del Y(\e{j\omega})}{\del \phi} &= \frac{-j\vecsym{\beta}^H \vec{v}(\psi) \e{-j\phi}}{(1 - \vecsym{\beta}^H \vec{v}(\psi) \vecsym{\alpha}^H \vec{v}(\psi)\e{-j\phi})^2}S(\e{j\omega})
\end{align*}
\end{footnotesize}

All the terms of $\vec{J}$ can be computed as:
\begin{subequations}
\begin{equation*}
	\resizebox{\hsize}{!}{$
	J_{\psi\psi} = \frac{1}{2\pi \sigma^2} \int_{-\frac{\omega_s}{2}}^{\frac{\omega_s}{2}} \left| \frac{\vecsym{\beta}^H (\vec{I} + \vec{v}(\psi)\vecsym{\beta}^H\vec{v}(\psi)\vecsym{\alpha}^H \e{-j\phi}) \vec{A}\vec{v}(\psi)}{(1 - \vecsym{\beta}^H \vec{v}(\psi) \vecsym{\alpha}^H \vec{v}(\psi)\e{-j\phi})^2} \right|^2 |S(\e{j\omega})|^2 \dif \omega
	$}
\end{equation*}

\begin{equation*}
\resizebox{\hsize}{!}{$
	J_{\phi\phi} = \frac{1}{2\pi \sigma^2} \int_{-\frac{\omega_s}{2}}^{\frac{\omega_s}{2}} \left| \frac{\vecsym{\beta}^H \vec{v}(\psi)}{(1 - \vecsym{\beta}^H \vec{v}(\psi) \vecsym{\alpha}^H \vec{v}(\psi)\e{-j\phi})^2}\right|^2 |S(\e{j\omega})|^2 \dif \omega
	$}
\end{equation*}

\begin{equation*}
\resizebox{\hsize}{!}{$
	J_{\psi\phi} = \Re\left\{\frac{j}{2\pi \sigma^2} \int_{-\frac{\omega_s}{2}}^{\frac{\omega_s}{2}} \frac{\vecsym{\beta}^H \left(\vec{I} + \vec{v}(\psi)\vecsym{\beta}^H\vec{v}(\psi)\vecsym{\alpha}^H \e{-j\phi}\right)\vec{A} \vec{v}(\psi) \vec{v}(\psi)^H \vecsym{\beta}}{(1 - \vecsym{\beta}^H \vec{v}(\psi) \vecsym{\alpha}^H \vec{v}(\psi)\e{-j\phi})^4}\right\} |S(\e{j\omega})|^2  \dif \omega
	$}
\end{equation*}
\end{subequations}
where $\vec{A} = \text{diag}(\begin{bmatrix} 0 &-j &-2j &\dots &-j(N-1)] \end{bmatrix})$.
Maximising this FIM results in the weights to be
\begin{align*}
	\vecsym{\beta} &= \frac{k\vec{v}(\psi)}{N}\e{-j(1-l)\phi},~\text{and} \\
	\vecsym{\alpha} &= \frac{\vec{v}(\psi)}{kN}\e{-jl\phi}, ~k \in \MB{C}-\{0\}, ~l \in [0,1].
\end{align*}
This expression is shown in \eqref{eq2:arrayfb} in Section~\ref{sec:feedbackbf}
without the parameter $\phi$.

\bibliography{IEEEabrv,bibly}

\begin{thebibliography}{10}

\bibitem{haykin2006cognitive}
Simon Haykin,
\newblock ``Cognitive radar: a way of the future,''
\newblock {\em IEEE signal processing magazine}, vol. 23, no. 1, pp. 30--40,
  2006.

\bibitem{li2014robust}
Yongzhe Li, Sergiy~A Vorobyov, and Aboulnasr Hassanien,
\newblock ``{Robust Beamforming for Jammers Suppression in MIMO Radar},''
\newblock in {\em 2014 IEEE Radar Conference}. IEEE, 2014, pp. 0629--0634.

\bibitem{liu2017robust}
Fan Liu, Christos Masouros, Ang Li, and Tharmalingam Ratnarajah,
\newblock ``{Robust MIMO Beamforming for Cellular and Radar Coexistence},''
\newblock {\em {IEEE} Wireless Commun. Lett.}, vol. 6, no. 3, pp. 374--377,
  2017.

\bibitem{van2002optimum}
Harry~L Van~Trees,
\newblock {\em {Optimum Array Processing: Part IV of Detection, Estimation, and
  Modulation Theory}},
\newblock John Wiley \& Sons, 2002.

\bibitem{vetterli2002sampling}
Martin Vetterli, Pina Marziliano, and Thierry Blu,
\newblock ``{Sampling Signals with Finite Rate of Innovation},''
\newblock {\em {IEEE} Trans. Signal Process.}, vol. 50, no. 6, pp. 1417--1428,
  2002.

\bibitem{sarkar1995using}
Tapan~K Sarkar and Odilon Pereira,
\newblock ``{Using the Matrix Pencil Method to Estimate the Parameters of a Sum
  of Complex Exponentials},''
\newblock {\em {IEEE} Antennas Propag. Mag.}, vol. 37, no. 1, pp. 48--55, 1995.

\bibitem{sathe2022automatic}
Prajakta Sathe and Amitabha Bhattacharya,
\newblock ``{Automatic Object Discrimination Based on Natural Resonant Features
  of Dielectric Coated Objects},''
\newblock {\em {IEEE} Trans. Antennas Propag.}, 2022.

\bibitem{venkitaraman2022annihilation}
Arun Venkitaraman and Pascal Frossard,
\newblock ``{Annihilation Filter Approach for Estimating Graph Dynamics from
  Diffusion Processes},''
\newblock in {\em ICASSP 2022-2022 IEEE International Conference on Acoustics,
  Speech and Signal Processing (ICASSP)}. IEEE, 2022, pp. 5583--5587.

\bibitem{1143830}
R.~Schmidt,
\newblock ``{Multiple Emitter Location and Signal Parameter Estimation},''
\newblock {\em {IEEE} Trans. Antennas Propag.}, vol. 34, no. 3, pp. 276--280,
  1986.

\bibitem{roy1989esprit}
Richard Roy and Thomas Kailath,
\newblock ``{ESPRIT-Estimation of Signal Parameters via Rotational Invariance
  Techniques},''
\newblock {\em {IEEE} Trans. Acoust., Speech, Signal Process.}, vol. 37, no. 7,
  pp. 984--995, 1989.

\bibitem{khan2008analysis}
ZI~Khan, M~MD Kamal, N~Hamzah, K~Othman, and NI~Khan,
\newblock ``{Analysis of Performance for Multiple Signal Classification (MUSIC)
  in Estimating Direction of Arrival},''
\newblock in {\em 2008 IEEE International RF and Microwave Conference}. IEEE,
  2008, pp. 524--529.

\bibitem{elbir2020deepmusic}
Ahmet~M Elbir,
\newblock ``{DeepMUSIC: Multiple Signal Classification via Deep Learning},''
\newblock {\em IEEE Sensors Letters}, vol. 4, no. 4, pp. 1--4, 2020.

\bibitem{hwang2008direction}
HK~Hwang, Zekeriya Aliyazicioglu, Marshall Grice, and Anatoly Yakovlev,
\newblock ``{Direction of Arrival Estimation using a Root-MUSIC Algorithm},''
\newblock in {\em Proceedings of the International MultiConference of Engineers
  and Computer Scientists}. Citeseer, 2008, vol.~2, pp. 19--21.

\bibitem{gao2005generalized}
Feifei Gao and Alex~B Gershman,
\newblock ``{A Generalized ESPRIT Approach to Direction-of-Arrival
  Estimation},''
\newblock {\em {IEEE} Signal Process. Lett.}, vol. 12, no. 3, pp. 254--257,
  2005.

\bibitem{steinwandt2017generalized}
Jens Steinwandt, Florian Roemer, and Martin Haardt,
\newblock ``{Generalized Least Squares for ESPRIT-Type Direction of Arrival
  Estimation},''
\newblock {\em {IEEE} Signal Process. Lett.}, vol. 24, no. 11, pp. 1681--1685,
  2017.

\bibitem{duan2005new}
Huiping Duan, Boon~Poh Ng, and Chong~Meng See,
\newblock ``{A New Broadband Beamformer Using IIR Filters},''
\newblock {\em {IEEE} Signal Process. Lett.}, vol. 12, no. 11, pp. 776--779,
  2005.

\bibitem{yan2006optimal}
Shefeng Yan,
\newblock ``{Optimal Design of FIR Beamformer With Frequency Invariant
  Patterns},''
\newblock {\em Applied Acoustics}, vol. 67, no. 6, pp. 511--528, 2006.

\bibitem{duan2007broadband}
Huiping Duan, Boon~Poh Ng, Chong Meng~Samson See, and Jun Fang,
\newblock ``{Broadband Beamforming using TDL-Form IIR Filters},''
\newblock {\em {IEEE} Trans. Signal Process.}, vol. 55, no. 3, pp. 990--1002,
  2007.

\bibitem{salvati2016use}
Daniele Salvati, Carlo Drioli, and Gian~Luca Foresti,
\newblock ``{On the use of Machine Learning in Microphone Array Beamforming for
  Far-Field Sound Source Localization},''
\newblock in {\em 2016 IEEE 26th International Workshop on Machine Learning for
  Signal Processing (MLSP)}. IEEE, 2016, pp. 1--6.

\bibitem{alkhateeb2018deep}
Ahmed Alkhateeb, Sam Alex, Paul Varkey, Ying Li, Qi~Qu, and Djordje Tujkovic,
\newblock ``{Deep Learning Coordinated Beamforming for Highly-Mobile Millimeter
  Wave Systems},''
\newblock {\em IEEE Access}, vol. 6, pp. 37328--37348, 2018.

\bibitem{stoica2002robust}
Petre Stoica, Zhisong Wang, and Jian Li,
\newblock ``{Robust Capon Beamforming},''
\newblock in {\em Conference Record of the Thirty-Sixth Asilomar Conference on
  Signals, Systems and Computers, 2002.} IEEE, 2002, vol.~1, pp. 876--880.

\bibitem{li2003robust}
Jian Li, Petre Stoica, and Zhisong Wang,
\newblock ``{On Robust Capon Beamforming and Diagonal Loading},''
\newblock {\em IEEE transactions on signal processing}, vol. 51, no. 7, pp.
  1702--1715, 2003.

\bibitem{zheng2019robust}
Zhi Zheng, Tong Yang, Di~Jiang, and Wen-Qin Wang,
\newblock ``{Robust and Efficient Adaptive Beamforming using Nested Subarray
  Principles},''
\newblock {\em IEEE Access}, vol. 8, pp. 4076--4085, 2019.

\bibitem{liu2022reduced}
Tuanning Liu, Wenqiang Liu, Yuanping Zhou, and Kesong Fan,
\newblock ``{Reduced-Dimension MVDR Beamformer Based on Sub-array
  Optimization},''
\newblock {\em IET Communications}, vol. 16, no. 18, pp. 2183--2192, 2022.

\bibitem{oppenheim2001discrete}
Alan~V Oppenheim, John~R Buck, and Ronald~W Schafer,
\newblock {\em {Discrete-Time Signal Processing. Vol. 2}},
\newblock Upper Saddle River, NJ: Prentice Hall, 2001.

\bibitem{antonion1993digital}
A~Antonion,
\newblock ``{Digital Filters: Analysis, Design and Application},''
\newblock {\em McGraw Hill, New York}, vol. 5, pp. 886896, 1993.

\bibitem{wen2013extending}
Fuxi Wen, Boon~Poh Ng, and Vinod~Veera Reddy,
\newblock ``{Extending the Concept of {IIR} Filtering to Array Processing using
  Approximate Spatial {IIR} Structure},''
\newblock {\em Multidimensional Systems and Signal Processing}, vol. 24, no. 1,
  pp. 157--179, 2013.

\bibitem{karo2020source}
Itay~Yehezkel Karo, Tsvi~G Dvorkind, and Israel Cohen,
\newblock ``{Source Localization with Feedback Beamforming},''
\newblock {\em {IEEE} Trans. Signal Process.}, vol. 69, pp. 631--640, 2020.

\bibitem{sandhu1985real}
GS~Sandhu and AV~Saylor,
\newblock ``{A Real-Time Statistical Radar Target Model},''
\newblock {\em IEEE Transactions on Aerospace and Electronic Systems}, , no. 4,
  pp. 490--507, 1985.

\end{thebibliography}
\bibliographystyle{IEEEbib}

\end{document}